\documentstyle[12pt,epsfig]{article}
\newcommand{\etal}{{\it et al.}}
\newcommand{\tk}{\mbox{$\tau^- \!\rightarrow\!\nu_\tau\, {\rm K}^-$}}
\newcommand{\kl}{\mbox{${\rm K}^- \!\rightarrow\!\bar{\nu}_l\, l^- (\gamma)$}}
\newcommand{\kmu}{\mbox{${\rm K}^- \!\rightarrow\!\bar{\nu}_\mu\, \mu^- (\gamma)$}}
\newcommand{\ke}{\mbox{${\rm K}^- \!\rightarrow\!\bar{\nu}_{\rm e}\, {\rm e}^- (\gamma)$}}
\newcommand{\rk}{\mbox{$f_{\rm K}/f_\pi$}}
\newcommand{\fpi}{\mbox{$f_\pi$}}
\newcommand{\btkwa}{\mbox{$B_{\tau\rightarrow\nu{\rm K}}^{\rm WA}$}}

\newcommand{\bklwa}{\mbox{$B_{{\rm K}\rightarrow \bar{\nu} l}^{\rm WA}$}}
\newcommand{\btkmssm}{\mbox{$B_{\tau\rightarrow\nu{\rm K}}^{\rm SM}$}}

\newcommand{\bklmssm}{\mbox{$B_{{\rm K}\rightarrow \bar{\nu} l}^{\rm SM}$}}
\newcommand{\vus}{\mbox{$|{\rm V}_{us}|$}}

\newcommand{\kpe}{\mbox{${\rm K}^- \!\rightarrow\! \pi^- \bar{\nu}_{\rm e}\, {\rm e}^-$}}
\newcommand{\rbeta}{\mbox{$\tan{\!\beta}/m_{\rm H}$}}
\newcommand{\rh}{\mbox{${\cal{R}}_{\rm H}$}}

\newcommand{\igev}{\mbox{GeV$^{-1}$}}

\begin{document}
\sloppy
\begin{titlepage}
\begin{flushright}
  hep-ex/0004022 \\  
  OPAL-IP082 \\
  April 26, 2000\\
% {\today}  \\
\end{flushright}
\bigskip\bigskip
    \begin{center}
     {\LARGE{\bf 
         Charged Higgs Mass Limits \\
         from the 
         $\tau^-\rightarrow \nu_\tau {\rm K}^-$ and 
         ${\rm K}^- \rightarrow \bar{\nu}_l l^- (\gamma)$ \\
         Branching Fractions
     }}
    \end{center}
    \begin{center} {\large S.\ Towers} \\
    {\sl Ottawa-Carleton Institute for Physics \\
         Carleton University \\
         Ottawa, Ontario K1S~5B6, Canada}\\
         {\tt smjt@physics.carleton.ca}
    \end{center}
    \begin{abstract}
From an analysis of the current world averages of 
the \kl\ and \tk\ branching fractions, we derive,
within the framework of type II Higgs doublets models such as
the Minimal Supersymmetric Extension of the Standard Model,
${\rbeta<0.21}$~\igev at a 90\% confidence limit.
\end{abstract}
\bigskip\bigskip
\begin{center}
{\large Submitted to Physics Letters {\bf B}}
\end{center}
\end{titlepage}

\section{Introduction}
The two Higgs doublet model represents one of the simplest
extensions to the minimal Standard Model, and is also of interest
in the context of supersymmetric models, which require the introduction
of at least two Higgs doublets to allow for spontaneous symmetry
breaking.  The Minimal Supersymmetric Extension of the Standard Model
(MSSM) is one of the more popular examples of such a model, and predicts
the existence of five Higgs bosons, three of which are neutral, and two of
which are charged (H$^+$ and H$^-$).  The ratio of the vacuum expectation
values of the two Higgs fields is an additional parameter in this model,
and is referred to as $\tan{\!\beta}$.
 
The most stringent previous
indirect limit on the mass of the charged Higgs
and $\tan{\!\beta}$ is the  
limit\footnote{Unless otherwise specified,
all limits presented in this note are the $90\%$
confidence limits.} 
on the ratio ${\rbeta<0.27}$~\igev\ 
obtained
from an analysis of ${B^+\rightarrow{\nu}_\tau\tau^+}$ decays collected
by the L3 experiment at LEP~\cite{bib:l3} \cite{bib:l3b}.
Previous analyses 
have also obtained ${\rbeta<0.67}$~\igev\ from studies of the
leptonic decays of the
tau lepton~\cite{bib:stahl1}, and ${m_{\rm H}>[244 + 63/(\tan{\!\beta})^{1.3}]}$ GeV
at a 95\% confidence limit
from an analysis of the ${b\rightarrow s\gamma}$ branching ratio~\cite{bib:alam}.  
Direct limits on the charged Higgs mass
have also been set by CDF, D0, and the four LEP 
experiments\cite{bib:cdf,bib:lep}.

In the work presented here, we explore possible charged Higgs effects in
\tk\ and \kl\ decays.
The branching
ratios for these decay modes are well measured, and the Standard
Model predictions are believed
to be well understood.  
 
\section{Decays of the Tau Lepton and Charged Kaon in the Standard 
Model}
The Standard Model prediction for the \tk\ branching ratio, including
${\cal{O}}(\alpha)$ electroweak corrections,
is given by \cite{bib:barish}
\begin{eqnarray}
B(\tau^- \rightarrow {\rm K}^- \nu_\tau) =
{{G_F^2 m_\tau^3 \tau_\tau} \over {16 \hbar \pi}} 
f_{\rm K}^2 |{\rm V}_{us}|^2
\left( 1 + {{2\alpha}\over{\pi}} \log{m_{\rm Z}/m_\tau} \right) 
\left( 1-m_{\rm K}^2/m_\tau^2 \right)^2 ,
\label{eqn:tau}
\end{eqnarray}
and
the 
prediction for the \kl\ branching ratio 
is \cite{bib:pdg}
\begin{eqnarray}
B({\rm K}^- \rightarrow l^- \bar{\nu}_l (\gamma)) =
{{G_F^2 m_{\rm K} m_l^2 \tau_{\rm K}} \over {8 \hbar \pi}} 
f_{\rm K}^2 |{\rm V}_{us}|^2
\left( 1 + {{2\alpha}\over{\pi}} \log{m_{\rm Z}/m_\rho} \right) 
\left( 1-m_l^2/m_{\rm K}^2 \right)^2 .
\label{eqn:kaon}
\end{eqnarray}
Equation~\ref{eqn:kaon} is commonly used to derive the kaon
decay constant, $f_{\rm K}$, from the experimental 
measurements of the \kmu\ branching ratio, but the 
constant can also
be obtained from lattice QCD calculations \cite{bib:QCD1}\cite{bib:QCD3}, 
or from other theoretical
calculations, such as those based upon the relativistic constituent
quark model \cite{bib:sant}\cite{bib:maris}.
%The value of the CKM matrix element, \vus, is most precisely derived from
%experimental measurements of the \kpe\ branching ratio~\cite{bib:vus}. 
 
\section{Charged Kaon and Tau Decays in the MSSM}
In theories where multiple Higgs doublets are responsible for 
spontaneous symmetry
breaking of the electroweak interactions, charged Higgs particles
arise as a natural consequence of the theory.  These charged
Higgs particles mediate charged current weak decays in the same manner
as the W$^\pm$, only differing in the Lorentz structure at the
decay vertex and in the coupling
strength of the boson to the fermions.  
The MSSM predicts the existence of two Higgs doublets,
where the first and second doublets couple to the
up- and down-type quarks, respectively, and where the strength
of the coupling to fermions is proportional to the fermion mass.
The Standard Model
branching ratios for the leptonic decays of the kaon and
the decay of a tau to a pseudoscalar kaon 
are both 
modified in the MSSM by a multiplicative factor \rh \cite{bib:hou}\cite{bib:abbott}
:
\begin{eqnarray}
\rh = \left( 1 - {{m_{\rm K}^2 \tan^2{\!\beta}}\over{m_{\rm H}^2}} \right)^2.
\label{eqn:rk}
\end{eqnarray}
We obtain the charged Higgs mass limits by comparing the experimental
world average
\tk\ and \kl\ branching fractions to the values predicted by
Equations~\ref{eqn:tau} and \ref{eqn:kaon}, multiplied by \rh
\footnote{It should be noted here that QCD and SUSY
radiative corrections to \rh\
are expected to be either small, or modify the factor in such a way
that the limit on \rbeta\ from this analysis would actually improve
if they were included \cite{bib:sola}.}.

\section{The Kaon Decay Constant}
We must ensure
that the value of $f_{\rm K}$ used in these studies is not
related to the measured \kl\ branching ratios, and therefore
independent from possible charged Higgs effects.
We thus use $f_{\rm K}$ derived  
from lattice QCD calculations;
to cancel many of the systematics associated with such calculations,
we use the calculated ratio \rk\,
multiplied by $f_\pi$ determined from experiment~\cite{bib:pdg}.
The values of \rk\ determined by two independent lattice calculations 
are shown in Table~\ref{tab:const}.  The value of $f_{\rm K}$ derived
from the average of these results
is $f_{\rm K} = 0.1552\pm0.0024$~GeV.
 
As a cross-check of the lattice calculational framework, 
we also examine the value of 
$f_{\rm K^\ast}/f_\rho$
determined by the same calculations.
These results are 
also shown in Table~\ref{tab:const}, and
are in agreement with each other, and also with the experimental
measurement obtained from the 
${\tau^-\!\rightarrow\! \nu_\tau {\rm K}^\ast}$ and 
${\tau^-\!\rightarrow\! \nu_\tau \rho}$ branching fractions 
\cite{bib:alephk}\cite{bib:randy},
${f_{{\rm K}^\ast}/f_\rho = 1.067 \pm 0.030}$.
 
The kaon decay constant can also be derived from calculations
based upon the relativistic quark model.  The author calculates, using
the formalism presented in reference \cite{bib:sant}, 
${f_{\rm K} = 0.1579\pm0.0044}$~GeV, where the uncertainty is due to
the uncertainties on the experimental values of inputs to the
calculation.   These inputs include the $\pi^0\rightarrow\gamma\gamma$
branching fraction, the charged pion decay constant, and the ${\rm K}_{{\rm e}3}$
form factor.  Unlike the studies presented in reference \cite{bib:sant},
the inputs do not include the experimental value of $f_{\rm K}$.
The author's calculation of $f_{\rm K}$ 
is in agreement with the lattice results, and
also with the result ${f_{\rm K} = 0.155}$~GeV presented in reference
\cite{bib:maris}. Reference \cite{bib:maris} also estimates that the
additional
model dependent uncertainties on $f_{\rm K}$ are on the order of
a few percent.  In these studies, we thus choose to use the more precise
lattice results, rather than this calculated value of $f_{\rm K}$.
 
\section{${\rm V}_{us}$}
The most
precise value of the CKM matrix element, ${\rm V}_{us}$, is
derived from experimental \kpe\ results \cite{bib:vus}, and
is relatively free from charged Higgs effects.
The author finds, with
calculations based on those presented in references
\cite{bib:hou2} and \cite{bib:belanger}, that  
the Standard Model \kpe\
branching ratio is multiplied in the MSSM by a factor 
which is generally much closer to
one than \rh,
and thus charged Higgs effects
on \vus\ are neglected here.

\section{Fitting Procedure}
To determine \rh\ we minimise the $\chi^2$:
\begin{eqnarray}
\chi^2 & = &
        \left( {  
                  {
                     B_{\tau\rightarrow\nu{\rm K}}^{\rm WA} 
               - \rh B_{\tau\rightarrow\nu{\rm K}}^{\rm SM}(\vec{X})
                  }
                 \over
                  {
                     \Delta B_{\tau\rightarrow\nu{\rm K}}^{\rm WA}
                  } 
               }
         \right)^2 
\nonumber \\
 & + &
        \left( { 
                 {
                     B_{{\rm K}\rightarrow \bar{\nu} \mu}^{\rm WA} 
               - \rh B_{{\rm K}\rightarrow \bar{\nu} \mu}^{\rm SM}(\vec{X})
                 }
                 \over
                 {
                     \Delta B_{{\rm K}\rightarrow \bar{\nu} \mu}^{\rm WA}
                 } 
               }
         \right)^2
%\nonumber \\
% & + &
+
        \left( { 
                 {
                     B_{{\rm K}\rightarrow \bar{\nu} {\rm e}}^{\rm WA} 
               - \rh B_{{\rm K}\rightarrow \bar{\nu} {\rm e}}^{\rm SM}(\vec{X})
                 }
                 \over
                 {
                     \Delta B_{{\rm K}\rightarrow \bar{\nu} {\rm e}}^{\rm WA}
                 } 
               }
         \right)^2
\nonumber \\
 & + &
 \sum_i \left( {
                 {X_i^{\rm WA} - X_i}
                 \over
                 {\Delta X_i^{\rm WA}}
               } \right)^2,
\label{eqn:chi}
\end{eqnarray}
where \btkwa\ and \bklwa\ refer to the world averages of the
\tk\ and kaon leptonic branching ratios, and where \btkmssm\ and \bklmssm\
are the Standard Model predictions for
the branching ratios derived from
Equations~\ref{eqn:tau} and \ref{eqn:kaon}, respectively.  Inputs to 
the predicted branching ratios,
such as \vus, the tau lifetime,
etc., are contained in the vector $\vec{X}$;                        
the vector $\vec{X}^{\rm WA}$
contains the world averages, while $\vec{X}$ contains the quantities
allowed to float in the fit. 
Note that this approach ensures that the uncertainty
on \rbeta\ includes, in
a natural way, the uncertainties on  
all of the quantities
input to the fit.  Unless otherwise specified, 
most of the inputs to the fit, such as lifetimes, and
the meson and lepton masses, are taken from reference~\cite{bib:pdg}. 
The least precise inputs to the fit are shown in Table~\ref{tab:input}.
 
The experimental world average branching ratios input to the
fit are shown in Table~\ref{tab:result}.  The kaon 
branching fractions
are taken from reference \cite{bib:pdg}, while the world
average of the \tk\ branching fraction is 
derived from results presented in references~\cite{bib:pdg} and
\cite{bib:aleph2}.
 
\section{Fit Results}
\begin{figure}[p]
   \begin{center}
     \mbox{ \epsfxsize=15cm
%           \epsffile[23 165 524 652]{eps/rh.ps} }
            \epsffile[23 165 524 652]{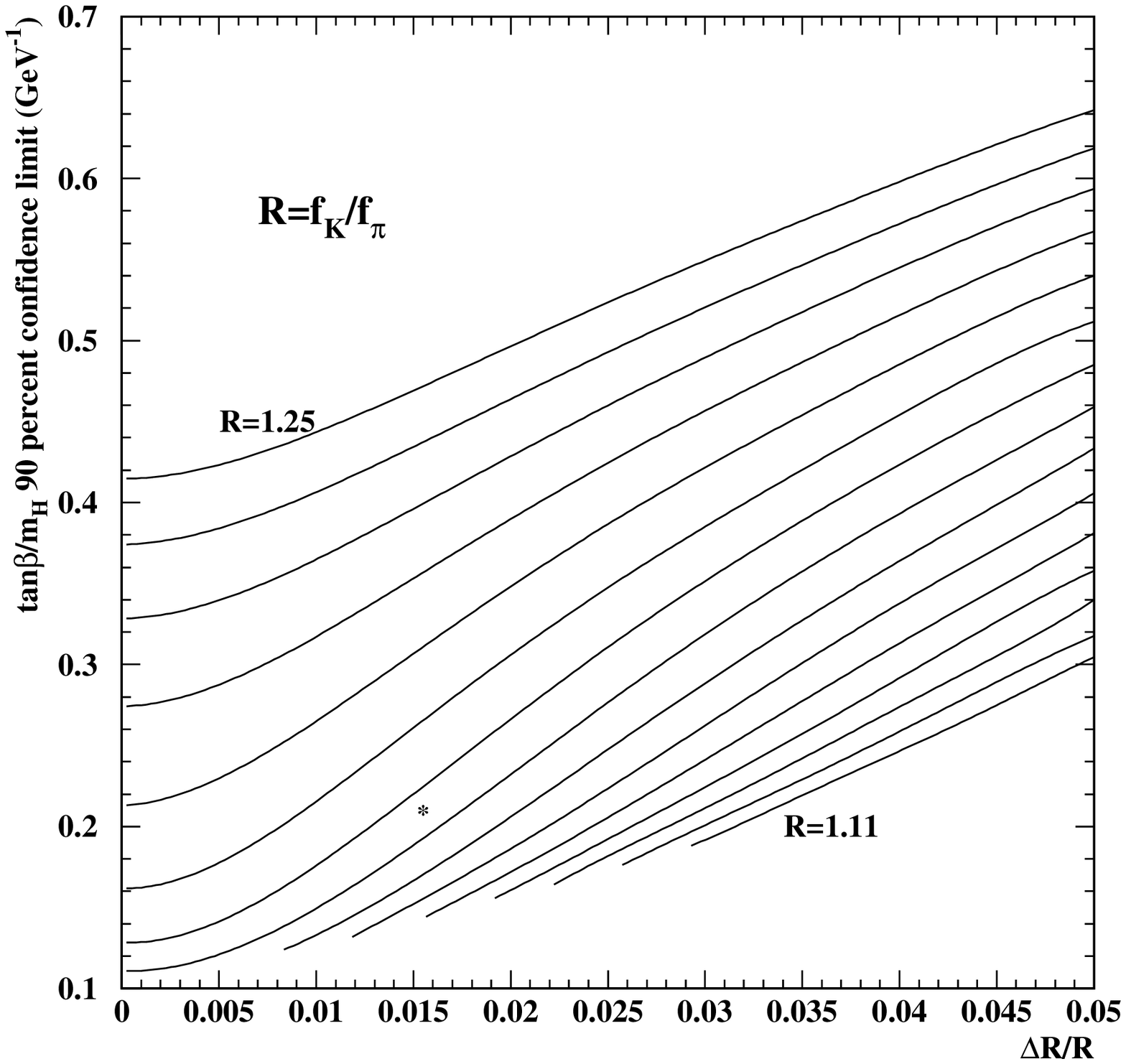} }
   \end{center}
 \caption[foo]{
 \label{fig:smass}
 The 90\% confidence limit on \rbeta\ as a function of
 the fit input $R=\rk$ and its associated relative uncertainty.
 The separate curves correspond to values of \rk\ from $1.11$ to $1.25$
 in increments of $0.01$. The $\ast$
 indicates the limit set using $\rk = 1.186\pm0.018$.
 }
 \end{figure}

The $\chi^2$ fit using Equation~\ref{eqn:chi} returns 
$\rh = 1.057^{+0.041}_{-0.039}$.  Using the method presented
in reference~\cite{bib:cousins} to account for the unphysical
region of $\rh>1$, a limit of 
${\rbeta<0.21}$~\igev\ is determined at
a 90\% confidence limit.  
The limit
as a function of the fit input value
of \rk, along with its associated uncertainty, 
is shown in Figure~\ref{fig:smass}.

The least precise outputs from the
fit are shown in Table~\ref{tab:input}, and Table~\ref{tab:result}
shows the central values of the MSSM predictions for the
branching ratios, as obtained from the fit results.
The linear correlations between \rh\ and the fit variables
\rk, \vus, and \fpi\ are
$-80\%$, $-55\%$, and $-15\%$, respectively.
 When the fit is repeated, holding the values of all inputs fixed
 to the values returned by the first fit, the 
 limit obtained is ${\rbeta<0.12}$~\igev.
 
Fitting only to the \kmu\ branching ratio yields ${\rbeta<0.21}$~\igev, while
fitting only to the \tk\ (or the \ke) branching ratio 
yields ${\rbeta<0.35}$~\igev\ (${\rbeta<0.45}$~\igev).
Thus the combined limit on \rbeta\ is dominated by the result
from the \kmu\
branching ratio.  However, the large tau samples expected to be collected
by B meson factories such as Babar, Belle, and CLEOIII 
will dramatically
improve the precision of the \tk\ branching ratio, and thus improve
the corresponding limit on \rbeta, potentially rivaling the result obtained from the
current value of the  
%By fitting for \rbeta\ using only the \tk\ channel, and a
%projected future value of the 
%branching ratio of ${0.7050\pm0.0005\%}$, the limit ${\rbeta<0.22}$~\igev\ is
%obtained, rivaling the result obtained from the current value of the
\kmu\ branching ratio.
 
In addition, future improvements in the precision of inputs to the fit
such as \vus\ and \rk\ will also improve the limit on \rbeta.  
It is possible
that the value of \vus\ extracted from the
$\tau^-\!\rightarrow\! \nu_\tau {\rm K}^\ast$ and
$\tau^-\!\rightarrow\! \nu_\tau \rho$ decays expected to be
collected by the
B factory experiments will, when combined with the value of
$f_{\rm K^\ast}/f_\rho$ from lattice calculations, rival the
precision of the current value of \vus, especially if the
precision of the lattice results also improves.
 
\section{Conclusions}
\begin{figure}[p]
   \begin{center}
     \mbox{ \epsfxsize=15cm
%           \epsffile[23 165 524 652]{eps/lim.ps} }
            \epsffile[23 165 524 652]{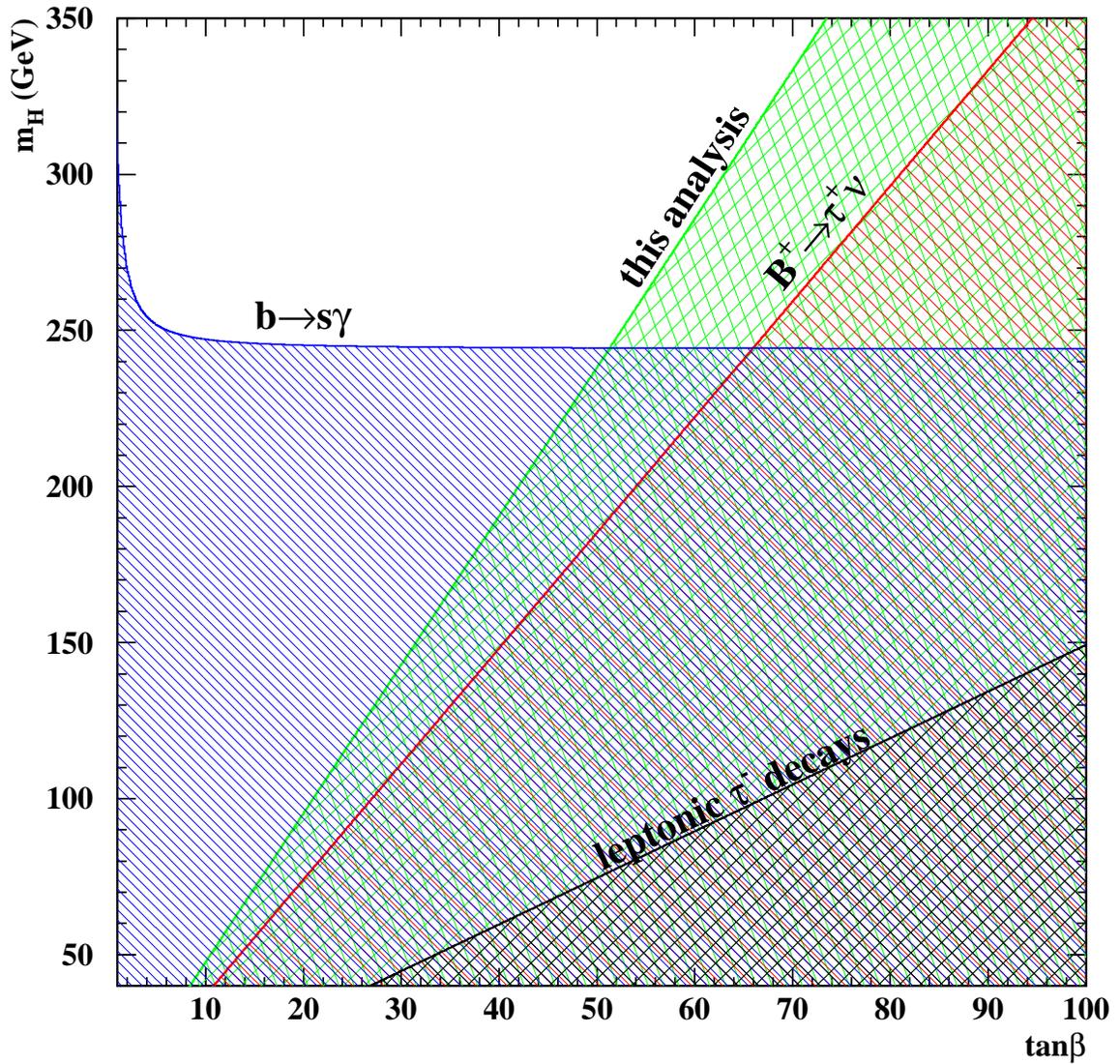} }
   \end{center}
 \caption[foo]{
 \label{fig:limit}
 Indirect limits on $\tan{\!\beta}$ and $m_{\rm H}$.  The limit from
 $b\rightarrow s \gamma$ studies is the 95\% confidence limit,
 while all other excluded regions
 reflect 90\% confidence limits.
 }
 \end{figure}

From a fit to the world average \kmu, \ke, and \tk\ branching
ratios we determine ${\rbeta<0.21}$~\igev\ at a 90\% confidence
limit. 
The area of the $m_{\rm H}$ versus $\tan{\beta}$ plane excluded by this,
and previous indirect limits, is shown in Figure~\ref{fig:limit}.
The limit is expected to improve once precise values
of the tau branching fractions are obtained by the various
B factory experiments.
 
\vspace*{0.9cm}
\par
\appendix
{\Large {\bf Acknowledgements}}
\vspace*{0.4cm}
\par
\noindent
I am indebted to P.\ Santorelli 
for many informative discussions regarding the determination of the
kaon decay constant in the Dyson-Schwinger formalism.
I would also like to thank W.S.\ Hou, H.H.\ Williams, S.\ Godfrey, 
and P.\ Maris
for their insightful commentary on various aspects of 
these studies. In addition, the financial support 
of the Ottawa-Carleton Institute
for Physics, and the hospitality of the High Energy Physics
group at the University of Pennsylvania over the period during which
this work was completed are both gratefully acknowledged.

\clearpage

%%%%%%%%%%%%%%%%%%%%%%%%%%%%%%%%%%%%%%%%%%%%%%%%%%%%%%%%%%%%%%%%%%%%%%%%%
%%%%%%%%%%%%%%%%%%%%%%%%%%%%%%%%%%%%%%%%%%%%%%%%%%%%%%%%%%%%%%%%%%%%%%%%%

%%%%%%%%%%%%%%%%%%%%%%%%%%%%%%%%%%%%%%%%%%%%%%%%%%%%%%%%%%%%%%%%%%%%%%%%%
%%%%%%%%%%%%%%%%%%%%%%%%%%%%%%%%%%%%%%%%%%%%%%%%%%%%%%%%%%%%%%%%%%%%%%%%%

\clearpage

%%%%%%%%%%%%%%%%%%%%%%%%%%%%%%%%%%%%%%%%%%%%%%%%%%%%%%%%%%%%%%%%%%%%%%%%%
%%%%%%%%%%%%%%%%%%%%%%%%%%%%%%%%%%%%%%%%%%%%%%%%%%%%%%%%%%%%%%%%%%%%%%%%%
\begin{table}[h]
\begin{center}
\begin{tabular}{|l|l|l|}
\hline
& $f_{\rm K}/f_{\pi}$ & $f_{{\rm K}^\ast}/f_\rho$ \\[1mm]
\hline
Reference \cite{bib:QCD1} & $1.20^{+0.03}_{-0.02}$ & $1.06^{+0.01}_{-0.02}$ \\
Reference \cite{bib:QCD3} & $1.13\pm0.04$          & $1.10\pm0.05$         \\ 
\hline
Average and $\chi^2$      & $1.186\pm0.018$  \hspace*{1.0cm} $\chi^2=2.5$      
                          & $1.061^{+0.010}_{-0.015}$  \hspace*{1.0cm} $\chi^2=0.9$   \\
%Experiment                & $1.223\pm0.012$        & $1.067\pm0.030$ \\
\hline
\end{tabular}
\end{center}
\vspace*{-0.5cm}
\caption[foo]{
\label{tab:const}
Meson decay constants from lattice QCD calculations.
}
\end{table}

%%%%%%%%%%%%%%%%%%%%%%%%%%%%%%%%%%%%%%%%%%%%%%%%%%%%%%%%%%%%%%%%%%%%%%%%%
%%%%%%%%%%%%%%%%%%%%%%%%%%%%%%%%%%%%%%%%%%%%%%%%%%%%%%%%%%%%%%%%%%%%%%%%%
\begin{table}[h]
\begin{center}
\begin{tabular}{|l|ll|l|}
\hline
& \multicolumn{2}{c|}{input value}
& value from fit \\           
\hline
\vus\                
                        & $0.2196\pm0.0023$ 
                        & \cite{bib:pdg} 
                        & $0.2196\pm0.0023$ \\
$f_\pi$                 & $0.13070\pm0.00039$ 
                        & \cite{bib:pdg} 
                        & $0.13070\pm0.00039$ \\
\rk                     & $1.186\pm0.018$ 
                        & see Table~\ref{tab:const}
                        & $1.186\pm0.018$ \\
\hline
\end{tabular}
\end{center}
\vspace*{-0.5cm}
\caption[foo]{
\label{tab:input}
Inputs to and outputs from the $\chi^2$ fit.
}
\end{table}

%%%%%%%%%%%%%%%%%%%%%%%%%%%%%%%%%%%%%%%%%%%%%%%%%%%%%%%%%%%%%%%%%%%%%%%%%
%%%%%%%%%%%%%%%%%%%%%%%%%%%%%%%%%%%%%%%%%%%%%%%%%%%%%%%%%%%%%%%%%%%%%%%%%
\begin{table}[h]
\begin{center}
\begin{tabular}{|l|l|l|}
\hline
& $B^{\rm WA}$ 
& $\rh B^{\rm SM}$ \\
\hline
\kmu\                   & $0.6406\pm0.0018$
                        & $0.6404$ \\
\ke\                    & $(1.55\pm0.07)\times10^{-5}$
                        & $0.1645\times10^{-5}$ \\
%\tk\                   & $0.00685\pm0.00022$
\tk\                    & $0.00694\pm0.00027$
                        & $0.00711$ \\
\hline
\end{tabular}
\end{center}
\vspace*{-0.5cm}
\caption[foo]{
\label{tab:result}
World average branching ratios input to the $\chi^2$ fit,
and the central values of the predicted branching ratios,
$\rh B^{\rm SM}$, returned by the fit.
}
\end{table}

\end{document}